\documentclass[epj]{webofc}
\usepackage[utf8]{inputenc}
\usepackage[varg]{txfonts}   
\usepackage{booktabs}
\usepackage{xcolor}
\definecolor{darkred}{rgb}{0.4,0.0,0.0}
\definecolor{darkgreen}{rgb}{0.0,0.4,0.0}
\definecolor{darkblue}{rgb}{0.0,0.0,0.4}
\usepackage[bookmarks,linktocpage,colorlinks,
    linkcolor = darkred,
    urlcolor  = darkblue,
    citecolor = darkgreen]{hyperref}
\usepackage{subfigure}
\wocname{EPJ Web of Conferences}
\woctitle{Lattice2017}
\begin{document}
%
\selectlanguage{english}
\title{%
An Alternative Lattice Field Theory Formulation Inspired by \\
Lattice Supersymmetry -Summary of the Formulation-
}
\author{%
\firstname{Alessandro} \lastname{D'Adda}\inst{1}\fnsep\thanks{1st speaker,\email{dadda@to.infn.it}} \and
\firstname{Noboru} \lastname{Kawamoto}\inst{2} \fnsep\thanks{2nd speaker,\email{kawamoto@particle.sci.hokudai.ac.jp}}\and
\firstname{Jun}  \lastname{Saito}\inst{3}
}
\institute{%
INFN Sezione di Torino, and
Dipartimento di Fisica Teorica,
Universita di Torino
I-10125 Torino, Italy 
\and
Department of Physics, Hokkaido University
Sapporo, 060-0810 Japan 
\and
Department of Human Science,
Obihiro University of Agriculture and Veterinary Medicine
Obihiro, 080-8555 Japan}
\abstract{%
We propose a lattice field theory formulation which overcomes some fundamental difficulties in  
realizing exact supersymmetry on the lattice. The Leibniz rule for the difference operator can 
be recovered by defining a new product on the lattice, the star product, and the chiral fermion 
species doublers degrees of 
freedom can be avoided consistently. This framework is general enough to formulate  
non-supersymmetric lattice field theory without chiral fermion problem. 
This lattice formulation has a nonlocal nature and is essentially equivalent to the corresponding 
continuum theory.  We can show that the locality of the star product is recovered exponentially 
in the continuum limit. Possible regularization procedures are proposed.The associativity of 
the product and the lattice translational invariance of the formulation will be discussed.   

}
\maketitle
\section{Introduction}\label{intro}
We have been asking ourselves the following question: "If we stick to keeping exact supersymmetry 
(SUSY) on the lattice,  
what kind of lattice formulation are we led to ?" There have been several proposals but they have not 
been completely successful as general 
formulations\cite{Dondi:1976tx, old-lattsusy,D-W-lattsusy, Kaplan2,Catterall1, Giedt3,Catterall3, 
Catterall4, Sugino,Kato-Sakamoto-So2,DKKN,Dadda-Kawamoto-Saito1,Damgaard-Matsuura, 
Kikukawa-Nakayama, Kadoh-Suzuki, DAdda:2010hbr, DAdda:2011drn}. 
There are several difficulties towards exact lattice supersymmetry which are of fundamental 
nature and intertwined with each other and thus not easy to solve at the same time. 
In particular it is difficult to keep SUSY and gauge invariance exactly at the 
same time for all super charges. We find a possible answer to our question by 
introducing on the lattice a new type of product, the star product which is nonlocal in nature but recovers 
locality exponentially in the continuum limit. Our lattice formulation based on the star product 
turns out to be equivalent 
to the continuum theory and thus it still needs regularization as we shall discuss in the end. 
The details of the formulation is given in \cite{DAdda:2017bzo}.   

\section{Difficulties of lattice SUSY and the possible solutions}\label{sec-1}
Let us consider the simplest possible SUSY algebra in one dimension. 
\begin{equation}
Q^2=i\partial 
\label{simple-SUSY-alg}
\end{equation}
There are two fundamental difficulties in formulating exact SUSY on the lattice. \\
i) Breakdown of the Leibniz rule for the difference operator. \\
ii) Chiral fermion species doubler problem.\\
In order to establish SUSY algebra on the lattice it is natural to replace the derivative 
operator by a difference operator. There is however an ambiguity in choosing which difference 
operator we introduce among forward, backward, symmetric difference operators. We claim 
Hermiticity requires symmetric difference operator as the lattice difference operator. 
In other words the momentum representation of lattice translation generator should be real:  
\begin{equation}
i\partial \psi(x) \rightarrow i\hat{\partial}\psi(x) \equiv i\frac{\psi(x+a)-\psi(x-a)}{2a} 
\rightarrow \frac{\sin ap}{a} \psi(p),   
\label{diff-op}
\end{equation}
where $a$ is the lattice constant and $p$ is the lattice momentum. 
The symmetric difference operator satisfies the following shifted Leibniz rule for the  
product of two fields: 
\begin{equation}
\hat{\partial}(\psi(x)\chi(x)) = \hat{\partial}\psi(x)\chi(x-a)+\psi(x+a)\hat{\partial}\chi(x) 
= \hat{\partial}\psi(x)\chi(x+a)+\psi(x-a)\hat{\partial}\chi(x). 
\label{latt-Leib}
\end{equation} 
Eq.(\ref{latt-Leib}) shows the breaking of the Leibniz rule for the difference operator and 
this is inconsistent with the super algebra (\ref{simple-SUSY-alg}) if Leibniz rule is 
assumed for $Q$ on the lattice. 

Naive fermion formulation on the lattice generates species doublers degree 
of freedom so the number of physical fermions will be increased compared to the number 
of bosons and this generates another source of SUSY breaking.  
In general this second difficulty ii) is considered to be a separate issue since chiral fermion 
problem for QCD is solved so that we may use the same method for the formulation of 
lattice SUSY. We, however, claim that exact SUSY on the lattice cannot be preserved 
unless the boson and fermion propagators have also on the lattice the 
simple relation ($D_{\hbox{bos}}=D_{\hbox{fer}}^2$) that they have in the continuum. 
Thus the second difficulty should be treated together with i). In fact we claim that the 
difficulties i) and ii) are fundamentally related. We need to solve these two difficulties at 
the same time. 

It is natural to consider $i\hat{\partial}$ defined in (\ref{diff-op}) as a generator of 
lattice translation. As one can see it generates a two steps lattice translation. It is natural 
to consider that physical states should be the eigenstates of the lattice translation 
generator. One can easily recognize that simple eigenstates of the translation generator 
(\ref{diff-op}) are:
\begin{equation}
 \psi=C,~~ ~~~ \psi'=C'(-1)^n,   
\end{equation}
where $C$ and $C'$ are constant and $x=na$ (n: integer) is lattice coordinate. The alternating 
sign state corresponds to a high frequency part in momentum space and thus corresponds 
to a species doubler state. We claim that we need to introduce single lattice translation 
generator since the lattice constant is the minimum unit of translation. It is then 
natural to introduce half lattice structure to accommodate alternating sign state where 
now lattice coordinate is $x=na/2$. It is then a natural question: "What is the half lattice 
translation generator ?" From the SUSY algebra (\ref{simple-SUSY-alg}) it is most 
natural to identify that the half lattice translation is lattice SUSY transformation. 
In fact we found that this algebraic and geometrical correspondence works for lattice 
superfield where the alternating sign $(-1)^{\frac{2x}{a}}$ interchanges bosons and fermions 
and plays a crucial role in the SUSY transformation\cite{DAdda:2010hbr}\cite{DAdda:2011drn}.    
 
For example in one dimension we introduce bosonic and fermionic lattice fields 
$\tilde{\Phi}(p)$ 
and $\tilde{\Psi}(p)$ in momentum representation to accommodate fermion species doublers 
and the bosonic counter parts to balance the degrees of freedom. 
We find the following $N=2$ SUSY transformation in one dimension:
\begin{eqnarray}
 Q_1 \tilde{\Phi}(p) =  i \cos \frac{a p}{4}  \tilde{\Psi}(p), ~~~~&
 Q_1 \tilde{\Psi}(p) = -4 i \sin \frac{a p}{4}   \tilde{\Phi}(p),\nonumber \\
Q_2 \tilde{\Phi}(p) =   \cos \frac{a p}{4}  \tilde{\Psi}\left(\frac{2\pi}{a}-p\right), ~~~~&
 Q_2 \tilde{\Psi}\left(\frac{2\pi}{a}-p\right) = 4  \sin \frac{a p}{4}   \tilde{\Phi}(p), 
\label{Qpb-f}
\end{eqnarray}
which satisfies the  $N=2$ SUSY algebra:
\begin{equation}
Q_1^2=Q_2^2=2\sin\frac{ap}{2}, ~~~~~~~~~~\{Q_1,Q_2\}=0,
\label{scharge-alg}
\end{equation}
where a dimensionless translation generator $2\sin\frac{ap}{2}$ is used here for simplicity. 
In this one dimensional $N=2$ model there are two fermions in the lattice field 
$\tilde{\Psi}(p)$; 
a particle state at $p=0$ and the species doubler at $p=\frac{2\pi}{a}$. In the boson 
sector we have corresponding counter part of fermions at each momentum region. 
In this way fermionic species doubler and their bosonic counter part constitute the 
super multiplet of $N=2$ SUSY algebra\cite{DAdda:2010hbr}\cite{DAdda:2011drn}. 
We consider this is one possible solution of the second problem: (A) The fermionic 
species doublers and the bosonic counterparts are identified as super partners. 

As we can see in (\ref{Qpb-f}) we can construct the exact $N=2$ SUSY algebra in the 
momentum space. In constructing the action we impose SUSY invariance up to the 
surface terms. Cancellation of the surface terms in the coordinate space can be 
translated into the total momentum conservation in the momentum space. In order 
to keep the algebraic structure of SUSY exact in the momentum space it is natural to 
identify the momentum representation of the translation generator as the conserved 
momentum. In other words we replace the lattice momentum conservation with the 
conservation of a single lattice translation generator:
\begin{equation}   
\delta(p_1+p_2+\cdots) \rightarrow \delta\left(\frac{2}{a}\sin\frac{ap_1}{2}+
\frac{2}{a}\sin\frac{ap_2}{2}+\cdots\right). 
\label{sine-mom-con}
\end{equation}
It turns out that this replacement solves the difficulty of the lattice Leibniz rule i). 
This type of replacement was first suggested by Dondi and Nicolai in their pioneering 
work of lattice SUSY in \cite{Dondi:1976tx}, where the conservation of a two step lattice 
translation generator $\frac{\sin ap}{a}$ was proposed. In our proposal we have 
introduced a half lattice structure to accommodate the geometric and algebraic 
correspondence of lattice SUSY transformation\cite{DAdda:2010hbr,DAdda:2011drn}.   

If we express the product of two fields $\Phi_1(x)\cdot\Phi_2(x)$ in the momentum 
space this leads to a convolution of two fields: 
\begin{equation}
\tilde{\Phi}_1(p)\cdot\tilde{\Phi}_2(p) =\frac{1}{2\pi}\int dp_1dp_2\delta(p-p_1-p_2)
 \tilde{\Phi}_1(p_1)\tilde{\Phi}_2(p_2).
\end{equation}
If we replace the momentum conservation with (\ref{sine-mom-con}) the product 
is changed from the normal product to a $\star$-product:
\begin{equation}
\tilde{\Phi}_1(p)\star\tilde{\Phi}_2(p) =\frac{1}{2\pi}\int dp_1dp_2\delta(\hat{p}-\hat{p}_1-\hat{p}_2)
 \tilde{\Phi}_1(p_1)\tilde{\Phi}_2(p_2), 
 \label{star-prod-mom}
\end{equation}
where $\hat{p_i}=\frac{2}{a}\sin\frac{ap_i}{2}$. 
It is important to realize that the conservation of the lattice translation generator  leads 
to the lattice Leibniz rule for the new $\star$-product: 
\begin{equation}
\hat{p}(\tilde{\Phi}_1(p)\star\tilde{\Phi}_2(p)) =\frac{1}{2\pi}\int dp_1dp_2\delta(\hat{p}-\hat{p}_1-\hat{p}_2)
 (\hat{p}_1\tilde{\Phi}_1(p_1)\tilde{\Phi}_2(p_2)+\tilde{\Phi}_1(p_1)\hat{p}_2\tilde{\Phi}_2(p_2)), 
 \label{leibniz-mom}
\end{equation}
where $\hat{p}_i=\frac{2}{a}\sin\frac{ap_i}{2}$ is nothing but a momentum representation of 
the difference operator. 

By parametrizing the delta function in (\ref{star-prod-mom}) we obtain the explicit presentation 
of the $\star$-product in coordinate space:
\begin{equation}
(\Phi_1\star\Phi_2)\left(\frac{na}{2}\right) =\sum_{n_1,n_2}K(n,n_1,n_2)\Phi_1\left(\frac{n_1a}{2}\right)
\Phi_2\left(\frac{n_2a}{2}\right),
\end{equation}
where 
\begin{equation}
K(n,n_1,n_2)=\int^\infty_{-\infty}d\lambda J_\Delta\left(\lambda,\frac{na}{2}\right)
J_\Delta\left(\lambda,\frac{n_1a}{2}\right)J_\Delta\left(\lambda,\frac{n_2a}{2}\right).
\end{equation}
Here $J_\Delta$ is Bessel function: 
\begin{equation}
J_\Delta\left(\lambda,\frac{na}{2}\right)=\frac{a}{2}\int^{\frac{4\pi}{a}}_0\frac{dp}{2\pi}
e^{-i\left(n\frac{ap}{2}-\lambda\frac{a\Delta(p)}{2}\right)},
\end{equation}
where $\Delta(p)=\hat{p}=\frac{2}{a}\sin\frac{ap}{2}$. As we can recognize the star product has 
a nonlocal nature. 

If we express eq. (\ref{leibniz-mom}) in coordinate space we can recognize that 
the single lattice translation difference operator satisfies exact Leibniz rule on the 
star product of fields, and thus the first difficulty i) is clearly solved. 
We can now construct actions of Wess-Zumino models for $N=2$ in one and two 
dimensions, which have exact lattice SUSY using the star product of 
fields\cite{DAdda:2010hbr}\cite{DAdda:2011drn}. In the one dimensional Wess-Zumino 
model species doublers of fermion and boson are identified as super multiplets of 
$N=2$ SUSY algebra. On the other hand in the two dimensional 
Wess-Zumino model\cite{DAdda:2011drn}  chiral conditions require the truncation 
of species doubler degrees of freedom:
\begin{equation}  
 \tilde{\Phi}_A(p_1,\cdots,p_j,\cdots,p_d)=\tilde{\Phi}_A(p_1,\cdots,\frac{2\pi}{a}-p_j,\cdots,p_d),
~~~(j=1,\cdots,d), 
\label{chiral-cond}
\end{equation}
where $A$ denotes both fermions and bosons and $d=2$ in two dimensions. 
In particular in the case of fermions we have identified the fermion with zero momentum 
and its species doubler $\left(p=\frac{2\pi}{a}\right)$. 
This is possible since species doubler has the same 
helicity as the one at the zero momentum due to the sign change of lattice momentum 
at $\frac{2\pi}{a}$: 
\begin{equation}
\frac{d\Delta(p)}{dp}|_{p=0}=\frac{d\Delta(\frac{2\pi}{a}-p)}{dp}|_{p=\frac{2\pi}{a}}.
\label{same-helicity}
\end{equation}
We identify this is the case (B) in contrast with (A): the identification of doublers as 
super partners. 

It is interesting to recognize at this stage that the truncation of the species doubler 
d.o.f. of fermions and of their bosonic counter parts can be systematically realized  
by imposing the condition (\ref{chiral-cond}) irrespective of chiral conditions.   
In other words even for a non SUSY formulation we can eliminate the extra 
d.o.f. of fields by introducing the half lattice structure and by truncating species doubler 
d.o.f. in each direction of dimensions. This can be well understood if we look into 
the coordinate version of the truncation condition (\ref{chiral-cond}):
\begin{equation} 
\Phi_A\left(\frac{n_1a}{2},\cdots,-\frac{n_ja}{2},\cdots,\frac{n_da}{2}\right)=
(-1)^{n_j}\Phi_A\left(\frac{n_1a}{2},\cdots,\frac{n_ja}{2},\cdots,\frac{n_da}{2}\right),
~~~(j=1,\cdots,d).
\label{chiral-cond-co}
\end{equation}
This condition also suggests that we only need to consider the first quadrant of 
coordinate space which has boundaries. This suggest naive breaking of lattice 
translational invariance of the formulation. 

There may also appear a worry that the lattice translational invariance is lost by 
changing lattice momentum conservation to $\Delta(p)$ momentum conservation 
in (\ref{star-prod-mom}). However the translational invariance of action is not 
lost but it is represented by infinitesimal transformations of the fields generated by 
$\Delta(p)$,
\begin{equation}
\delta_{\epsilon} \tilde{\varphi}_A(p) = \epsilon^{\mu} \Delta(p_\mu)\tilde{\varphi}_A(p), 
\label{latticetranslation} 
\end{equation}
where $\epsilon_\mu$ is infinitesimal parameter. 

We can thus conclude this section by remarking that  
the solution for the difficulties of lattice 
SUSY formulation i) and ii) have been obtained. The identification of the conserved 
momentum with the single lattice difference operator solves the Leibniz rule 
problem with the introduction of a new star product. The introduction of the half lattice 
structure accompanied by the truncation condition (\ref{chiral-cond}) solves 
the second problem.  

\section{Breakdown of associativity and its recovery in the star product}\label{sec-2}

In order to formulate exact lattice SUSY it is enough to solve the difficulties 
i) and ii). In fact we found a formulation of the $N=2$ Wess-Zumino models in one 
and two dimensions which have exact SUSY on the 
lattice\cite{DAdda:2010hbr,DAdda:2011drn}. 
It has been also confirmed that SUSY on 
the lattice is preserved even at the quantum level\cite{Asaka:2016cxm}. 
It has, however, been recognized that associativity for the $\star$-product is broken:
\begin{eqnarray}
&&(\tilde{\Phi}_1\star (\tilde{\Phi}_2\star \tilde{\Phi}_3))(p) \nonumber \\
&&=\int dp_{23}  \int dp_1\int dp_2 \int dp_3  
 \tilde{\tilde{\Phi}}_1(p_1) \tilde{\tilde{\Phi}}_2(p_2) \tilde{\tilde{\Phi}}_3(p_3) 
\delta(\hat{p}_{23}-\hat{p}_2-\hat{p}_3) 
\delta(\hat{p} -\hat{p}_1-\hat{p}_{23}) \nonumber \\
&&\neq \int dp_{12}  \int dp_1\int dp_2 \int dp_3   
\tilde{\tilde{\Phi}}_1(p_1) \tilde{\tilde{\Phi}}_2(p_2) \tilde{\tilde{\Phi}}_3(p_3) 
\delta(\hat{p}_{12}-\hat{p}_1-\hat{p}_2) 
\delta(\hat{p} -\hat{p}_3-\hat{p}_{12}) \nonumber \\
&&=((\tilde{\Phi}_1\star \tilde{\Phi}_2)\star \tilde{\Phi}_3)(p)
\label{non-assoc-1}
\end{eqnarray}
where  $\hat{p}_i=\frac{2}{a}\sin\frac{ap_i}{2}$.
This is because there is a region of phase space where the following domains 
(1) and (2) do not overlap: 
\begin{eqnarray} 
(1)&&~~~|\hat{p}_i|<\frac{2}{a}, ~~|\hat{p}_2+\hat{p}_3|<\frac{2}{a}, 
~~|\hat{p}_1+\hat{p}_2+\hat{p}_3|<\frac{2}{a} \nonumber \\
(2)&&~~~|\hat{p}_i|<\frac{2}{a}, ~~|\hat{p}_1+\hat{p}_2|<\frac{2}{a}, 
~~|\hat{p}_1+\hat{p}_2+\hat{p}_3|<\frac{2}{a}.
\label{non-asso-phase-space}
\end{eqnarray}

\begin{figure}
\begin{center}
\mbox{\includegraphics*[width=6.cm]{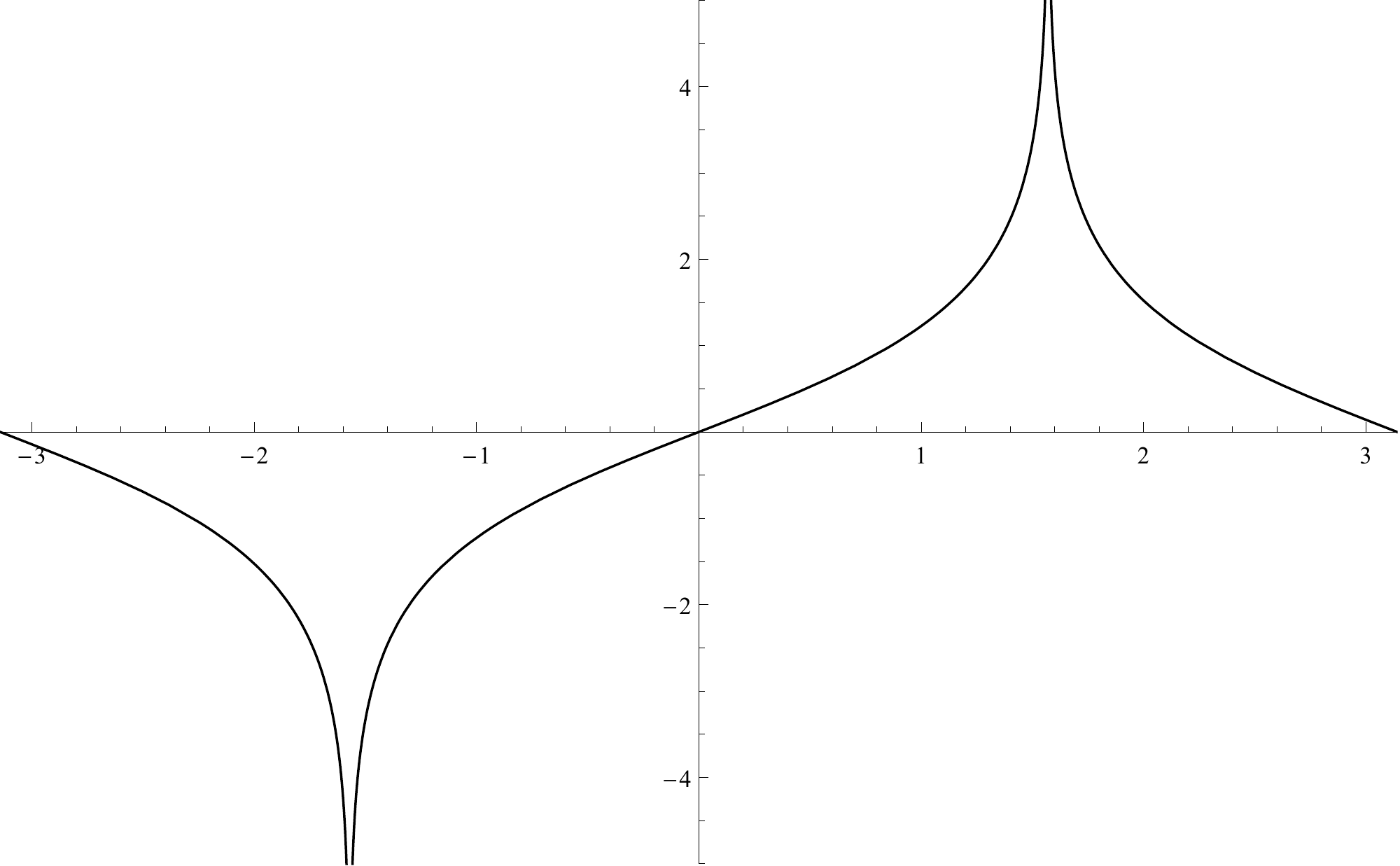}}
\caption{Plot of the Inverse Gudermannian function $gd^{-1}(x)$ in the fundamental interval $(-\pi,\pi)$}
\label{guderm}
\end{center}
\end{figure}

It is interesting to realize that the breakdown of associativity does not affect  
exact SUSY invariance on the lattice for non-gauge Wess-Zumino models. 
On the other hand the breakdown of the associativity for the $\star$-product makes it 
difficult to extend this formulation to gauge theories since 
associativity is crucial for the gauge invariance proof as we can see 
in the following example:
\begin{align}
\Phi^{\dagger}(x) \star\Phi(x) &\rightarrow (  \Phi^{\dagger}(x) \star 
e^{-i \alpha(x)}) \star  (   e^{i \alpha(x)}\star  \Phi(x))  \nonumber \\
&\neq  \Phi^{\dagger}(x) \star( e^{-i \alpha(x)} \star  e^{i \alpha(x)})\star  
\Phi(x) = \Phi^{\dagger}(x) \star \Phi(x).
\end{align}
In order to recover the associativity for the $\star$-product we notice that if 
momenta $\hat{p}_i$ does not have upper limit in (\ref{non-assoc-1}); then $\frac{2}{a}$ 
is replaced by $\infty$ in (\ref{non-asso-phase-space}) and 
associativity is recovered. For this reason we 
tried to find an alternative lattice translation generator $\Delta(p)$  with the following 
properties:
\begin{eqnarray}  
&&1)~ \Delta(-p)=-\Delta(p), \nonumber\\
&&2)~\frac{a}{2}\Delta(p)=\frac{ap}{2} + O\left(\left(\frac{ap}{2}\right)^3\right), 
\nonumber\\
&&3) ~\Delta(p) ~\hbox{has to cover twice the whole real axis as}~ p ~
\hbox{goes through the} ~\frac{4\pi}{a} ~\hbox{period.} \nonumber \\
&&4) ~\Delta(p) = \Delta\left(\frac{2\pi}{a}-p\right), ~~~
\lim_{p\rightarrow \pm\frac{\pi}{a}}\Delta(p)=\pm\infty. 
\end{eqnarray}
We found an almost unique solution: 
\begin{equation}
\hat{p}=\Delta_G(p)=\frac{1}{a}\log\frac{1+\sin\frac{ap}{2}}{1-\sin\frac{ap}{2}}.
\label{Delta-G}
\end{equation}
This is the inverse Gudermannian function and it has the following nice expansion:
\begin{equation} \frac{a}{2} \Delta_G(p) = 2 \left[  \sin \frac{ap}{2} -  \frac{1}{3}  
\sin \frac{3ap}{2}+  \frac{1}{5}  \sin \frac{5ap}{2}-  \frac{1}{7}  \sin \frac{7ap}{2}+\cdots \label{pwex2}\right], 
\end{equation}
which in coordinate representation is a sum over all odd half integer multiple of difference operators:\begin{equation}\Delta_G\Phi(x) = \frac{2}{a} \sum_{k=1}^{\infty}  \frac{(-1)^{k+1}}{2k-1}  \left[ \Phi\left(x+\frac{(2k-1)a}{2}\right) -  \Phi\left(x-\frac{(2k-1)a}{2}\right)\right],\label{deltacoord} 
\end{equation}
and so it is intrinsically nonlocal in nature. 
In fig.\ref {guderm} we show plot of the inverse Gudermannian function 
\begin{equation}
gd^{-1}(x) = \frac{1}{2} \log \frac{1+ \sin x}{1- \sin x}, \label{gd} 
\end{equation}
As we can see from the figure there is a species doubler at $x=\frac{ap}{2}=\pi$. 

It is important to note that the condition 4) is crucial to keep the symmetry of 
the whole action under the equivalence of (\ref{chiral-cond}) together with the 
condition (\ref{same-helicity}). In conclusion we have found a lattice field theory formulation which 
avoids all the difficulties to fulfill exact lattice SUSY: \\
i) Difference operator satisfies exact Leibniz rule on $\star$-product. \\
ii) No chiral fermion doublers by truncation of the doublers d.o.f. \\
iii)  Associativity is satisfied for the $\star$-product. \\
It turns out that this nonlocal lattice field theory formulation is equivalent to the corresponding continuum theory. How comes that this is possible ?

Let us consider the normal product of two fields $\tilde{\Phi}_1(p)$ and 
$\tilde{\Phi}_2(p)$ which is expressed by a convolution in the momentum space:
\begin{equation}
\widetilde{\Phi_1 \star \Phi_2 }(\hat{p}) = \frac{2}{\pi}  \int_{\Delta_G(-\frac{\pi}{a})}^{\Delta_G(\frac{\pi}{a})} d\hat{p}_1 d\hat{p}_2  \tilde{\Phi}_1(\hat{p}_1)\tilde{\Phi}_2(\hat{p}_2) \delta \left( \hat{p} - \hat{p}_1 - \hat{p}_2 \right), \label{contpr} 
\end{equation}
where $\hat{p}_i=\Delta_G(p_i)$ and $\Delta_G(\pm\frac{\pi}{a})=\pm\infty$. 
The $\star$-product at this stage is the same as the normal product as it can 
be seen by simply changing the integration variables:
\begin{eqnarray}
f(p)\frac{1}{f(p)}\frac{d\hat{p}}{dp}\widetilde{\Phi_1 \star \Phi_2 }(\hat{p})&=&
\frac{d\hat{p}}{dp} \frac{2}{\pi}  \int_{-\frac{\pi}{a}}^{\frac{\pi}{a}} 
dp_1dp_2 f(p_1)\nonumber \\
& &\frac{1}{f(p_1)}\frac{d\hat{p_1}}{dp_1}\tilde{\Phi}_1(\hat{p}_1)
f(p_2)\frac{1}{f(p_2)}\frac{d\hat{p_2}}{dp_2}\tilde{\Phi}_2(\hat{p}_2)
\delta \left( \hat{p} - \hat{p}_1 - \hat{p}_2 \right).
\label{equivalent-conv}
\end{eqnarray}
This relation can be replaced by the following equivalent relation: 
\begin{equation}
f(p) ~\widetilde{\varphi_1 \star \varphi_2 }(p) =\frac{2}{\pi}  \int_{-\frac{\pi}{a}}^{\frac{\pi}{a}} dp_1 dp_2  \frac{d\Delta_G(p)}{dp} f(p_1) \tilde{\varphi}_1(p_1) ~f(p_2) \tilde{\varphi}_2(p_2) \delta\left( \Delta_G(p) -\Delta_G(p_1) - \Delta_G(p_2) \right), 
\label{assstarprodpr} 
\end{equation}
where we identify the lattice wave function as:
\begin{equation}
  \tilde{\varphi}_i(p_i)= \frac{1}{f(p_i)}\frac{d\Delta_G(p_i)}{dp_i}  \tilde{\Phi}_i(\Delta_G(p_i)),~~~~~~~~~~~~~-\frac{\pi}{a}\leq p_i \leq \frac{\pi}{a}.
  \label{Phivarphi} 
\end{equation}
with $\hat{p}_i=\Delta_G(p_i)$. As one can see from (\ref{Phivarphi}) the lattice field 
$\tilde{\varphi}(p_i)$ is defined up to a renormalization factor 
$f(p_i)$. The factor $\frac{d\Delta_G(p_i)}{dp_i}=\frac{1}{{\cos\frac{ap}{2}}}$ 
plays a role of compensating the argument of delta function 
linear in $p$: $\delta(p-\cdots)$. 

The fact that the lattice and the continuum fields may be related by 
eq.(\ref{Phivarphi}) and that under this correspondence the field product of the 
continuum theory becomes the star product (\ref{assstarprodpr}) on the lattice 
is highly non trivial. It naturally raises the question: "Can the continuum and the 
lattice countable infinity have the same number of degrees of freedom ?" 
In other words: Can the continuum to lattice wave function relation be inverted ? 
We investigated this question in the coordinate space. It turned out that it is possible 
to make the correspondence invertible with a smooth continuum limit 
if we choose the function $f(p)$ as:
\begin{equation}  
f(p)=\sqrt{\frac{d\Delta_G}{dp}}=\frac{1}{\sqrt{\cos\frac{ap}{2}}}. 
\label{fp}
\end{equation}
The details of the arguments can be found in \cite{DAdda:2017bzo}.
The $\star$-product of lattice fields is then defined as:
\begin{eqnarray}
\sqrt{\cos\frac{ap}{2}}\widetilde{\varphi_1\star\varphi_2}(p) =
\frac{2}{\pi} \int^{\frac{\pi}{a}}_{-\frac{\pi}{a}} \frac{dp_1}{\sqrt{\cos\frac{ap_1}{2}}}
\frac{dp_2}{\sqrt{\cos\frac{ap_2}{2}}}\tilde{\varphi}_1(p_1)\tilde{\varphi}_2(p_2) 
\int d\xi e^{i\xi(\Delta_G(p)-\Delta_G(p_1)-\Delta_G(p_2))}.
\label{star-prod-G}
\end{eqnarray}
The coordinate representation of the $\star$-product is given by 
\begin{equation}
(\varphi_1\star\varphi_2)^{(0)}(x_n) = \frac{a^2}{4} \sum_{n_1,n_2} K_{n,n_1,n_2} \varphi^{(0)}_1(x_{n_1})\varphi^{(0)}_2(x_{n_2}),  \label{asprsymm} 
\end{equation}
where $(0)$ denotes the coordinate fields corresponding after the species doublers 
truncation. 
The kernel $K_{n,n_1,n_2}$ is completely symmetric in the three indices and is given by
\begin{equation}
K_{n,n_1,n_2} = \int_{-\infty}^{+\infty} d\xi ~J^{(0)}_{\Delta_G}(\xi,x_n)~
J^{(0)}_{\Delta_G}(\xi,x_{n_1})~J^{(0)}_{\Delta_G}(\xi,x_{n_2})
,\label{symmker} 
\end{equation}
with
\begin{equation}
J^{(0)}_{\Delta_G}(\xi,x_n) = \frac{1}{2\pi} \int_{-\frac{\pi}{a}}^{\frac{\pi}{a}}
\frac{dp}{\sqrt{\cos\frac{ap}{2}}} e^{ix_n p -i\xi \Delta_G(p)},  \label{JDeltaprime} 
\end{equation}
where $x_n=\frac{na}{2}$.

Lattice to continuum and continuum to lattice transformation of fields in coordinate 
space are given by
\begin{eqnarray}
\hspace{2cm}~~~~~~~~~~~~~~~~~~~~
\Phi(\xi) &=& \frac{a}{2} \sum_n~J^{(0)}_{\Delta_G}(\xi,x_n) \varphi_0(x_n), 
\nonumber \\
\hspace{2cm}~~~~~~~~~~~~~~~~~~~~
\varphi_0(x_n) &=& \int_{-\infty}^{\infty} d\xi~\bar{J}^{(0)}_{\Delta_G}(x_n,\xi)~
\Phi(\xi),   \label{varphi00}
\end{eqnarray}
where 
\begin{equation}
\bar{J}_{\Delta_G}^{(0)}(\eta,\xi) = J_{\Delta_G}^{(0)}(\xi,\eta) = 
\frac{N}{2\pi} \int_{-\frac{\pi}{2}}^{\frac{\pi}{2}} \frac{d\theta}{\sqrt{\cos(\theta)}}~
e^{iN\left(\xi gd^{-1}(\theta) - \eta \theta \right) }, \label{J0Deltasq} 
\end{equation}
with $N=\frac{2}{a}$ and $gd^{-1}(\theta)$ is given in (\ref{gd}) 
and we have also replaced $x_n=\frac{na}{2}$ with the continuum variable $\eta$:
\begin{equation}
x_n=\frac{n}{N} \rightarrow \eta ~~~(n\rightarrow \infty,~ N\rightarrow \infty).
\end{equation}
It is interesting to recognize that the arguments $\eta$ and $\xi$ are 
interchanged for $\bar{J}^{(0)}_{\Delta_G}$ and $J^{(0)}_{\Delta_G}$. 
This is similar to the Fourier transform and the inverse Fourier transform 
for $e^{ipx}$. 

\begin{figure}
\begin{center}
\mbox{\includegraphics*[width=6.cm]{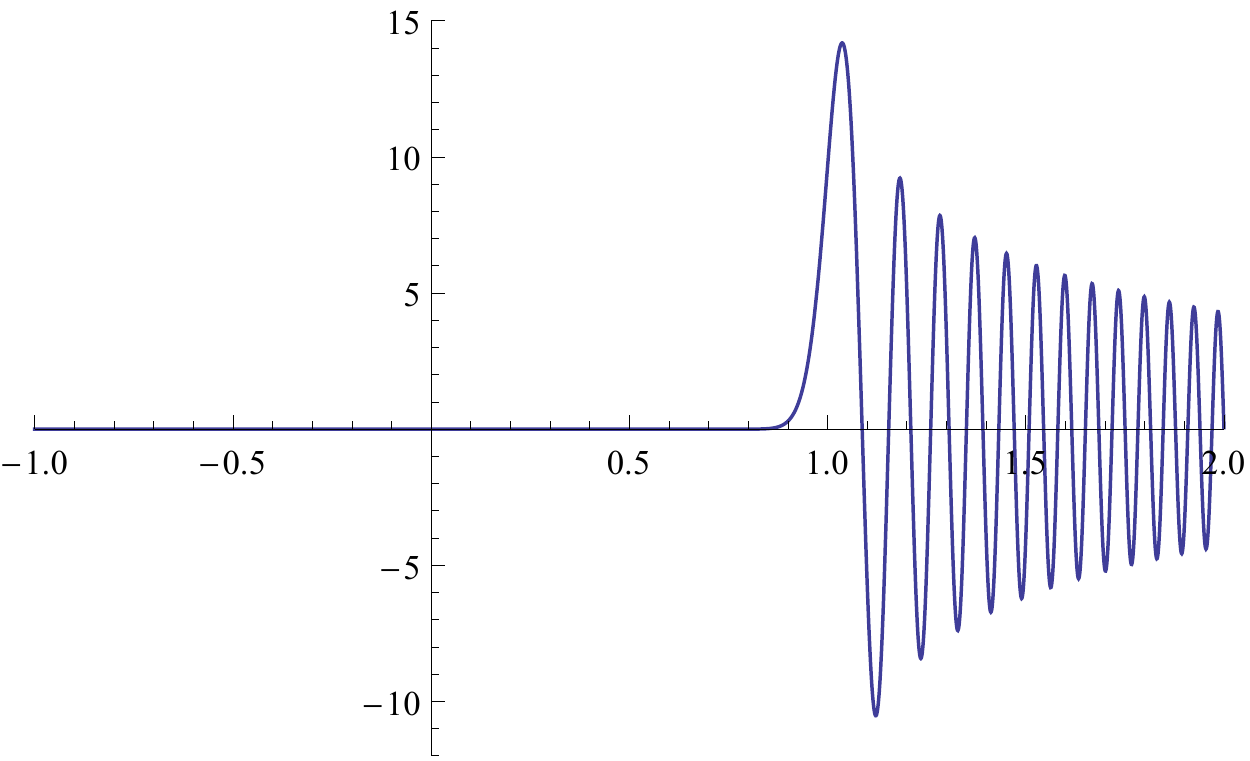}}
\caption{Plot of  $J_{\Delta_G}^{(0)}( \xi=1, \eta)$ versus $\eta$ at $N=100$.}
\label{Jplot3}
\end{center}
\end{figure}

As we can see again the $\star$-product defined in (\ref{asprsymm}) is 
nonlocal in nature. The first question here is: "Is the locality of the 
$\star$-product recovered in the continuum limit ?" The answer is 
"yes". We can show:     
\begin{equation}
\lim_{N\rightarrow \infty}J^{(0)}_{\Delta_G}(\xi,\eta) =\delta(\xi-\eta). 
\end{equation}
See the details in \cite{DAdda:2017bzo}. This assures the recovery of locality 
for the $\star$-product (\ref{asprsymm}) in the continuum limit. 
The next question is  "How much local is the $\star$-product ?" 
Here we show the explicit functional dependence of 
$J^{(0)}_{\Delta_G}(\xi=1,\eta)$ in fig.\ref{Jplot3}.
As we can see from the figure, around $\eta=1$  there is a peak, $\eta<1$ 
it decreases exponentially,  $\eta>1$ it oscillates violently as an analytic continuation 
of exponential damping and thus plays 
a role of delta function. In fact damping nature below and above $\eta=1$ 
is thus exponential in nature. We can observe that the nonlocal $\star$-product 
is effectively local in the continuum limit. 

We can now construct  a continuum equivalent lattice field theory as follows:\\
(1) Write down the momentum representation of a continuum action.  \\
(2) Replace the continuum derivative operator by $\Delta_G(p)$ and 
the lattice momentum conservation by the derivative operators conservation 
of $\Delta_G(p)$. \\
(3) Replace the continuum wave function by the lattice wave function as: 
\begin{equation}
\tilde{\varphi}_A(p_j)=\sqrt{\prod_j\frac{d\Delta_G(p_j)}{dp_j}}\tilde{\Phi}_A(\hat{p}_j) 
=\frac{1}{\sqrt{\prod_j\cos\frac{ap_j}{2}}}\tilde{\Phi}_A(\hat{p}_j). 
\label{cont-lat-trans}
\end{equation}
(4) Identify all the fields as given in eq. (\ref{chiral-cond}) and consider only 
the first quadrant of coordinate space. 
We then obtain a continuum equivalent lattice action which has all the symmetries  
of the corresponding continuum theory and no chiral fermion problem. 
The $\star$-product is now associative. 

All these replacements together 
with the substitution of (\ref{cont-lat-trans}) can be summarized in the 
following transformation from a continuum action to the corresponding lattice action: 
\begin{equation}
 e^{-\hat{S}_{\Delta_G}(\tilde{\varphi})} =  \int \mathcal{D}\tilde{\Phi}_A \prod_\mu \prod_{p_\mu=-\frac{\pi}{a}}^{\frac{\pi}{a}}\prod_A\delta\left( \tilde{\Phi}_A(\Delta_G(p)) - 
\sqrt{\prod_j\cos\frac{ap_j}{2}} ~ \tilde{\varphi}_A(p) \right) e^{-S_{cl}(\tilde{\Phi})}. \label{effect3} 
\end{equation}
It is interesting to recognize that the transformation (\ref{cont-lat-trans}) 
can be identified as a momentum representation of block spin type transformation. 
A typical block spin type 
transformation from continuum to lattice can be given in the coordinate space as:   
\begin{equation} 
\varphi_A(n) = \int dx f(nl-x) \Phi_A(x), \label{bltr} 
\end{equation}
where $f(nl-x)$ specifies blocking type. In this transformation the symmetries of the original action 
are not necessarily preserved. Criteria for the symmetries to be maintained on the lattice, 
at least in the Ginsparg-Wilson sense,  
were investigated in \cite{Bergner-Bruckmann}. In our case, namely the blocking 
transformation (\ref{effect3})  
the symmetries of original theory are preserved in the lattice action. 

There is, however, a crucial feature of this formulation namely that the lattice formulation 
is not yet regularized and the lattice constant is not playing the role of regulator. 
Nevertheless we have obtained a lattice field theory formulation which has the same  
lattice symmetries as the corresponding continuum theory have. 
We, however, need to regularize this lattice theory. 

\section{Regularization and renormalization in the new lattice}\label{sec-3}
As a possible regularization we introduce regularization parameter $\hat{z}$ 
into the regularized derivative operator as:
\begin{equation}
\Delta^{(\hat{z})}_G(p) = \frac{1}{a\hat{z}} 
\log\frac{1+\hat{z}\sin \frac{ap}{2}}{1-\hat{z}\sin \frac{ap}{2}},
\label{regguder} 
\end{equation} 
which extrapolates between two typical conserved momenta:
\begin{equation}  
\lim_{\hat{z}\rightarrow 0}\Delta^{(\hat{z})}_G(p)=\frac{2}{a}\sin \frac{ap}{2},~~~
\Delta^{(\hat{z}=1)}_G(p)=\Delta_G(p).
\end{equation}
For $\hat{z}<1$ the regularized derivative operator $\Delta^{(\hat{z})}_G(p)$ is bounded by
\begin{equation}
|\Delta^{(\hat{z})}_G(p)| \leq \frac{1}{a\hat{z}} \log\frac{1+\hat{z}}{1-\hat{z}}
=\hat{p}^{\hbox{(cutoff)}}, \label{bound} 
\end{equation}
and thus plays a role of cut off momentum. 

\subsection{Wess-Zumino model in four dimensions}\label{sec4-1}
Let us consider as an example, the 4 dimensional Wess-Zumino model. The kinetic terms 
of the action are given by:
\begin{equation}
S = \int d^4x \left( i \Psi^\dagger \bar{\sigma}^\mu\partial_\mu\Psi - \partial^\mu\Phi^\star\partial_\mu\Phi \right), \label{4dWSmodel} 
\end{equation} 
where $(\bar{\sigma}^\mu)=(\sigma^0,-{\sigma}^i)$. 
The action (\ref{4dWSmodel}) is invariant on shell under the supersymmetry transformations:
\begin{eqnarray}
&\delta_\epsilon \Phi = \epsilon \Psi,~~~~~~~~~~~~~~&\delta_\epsilon\Phi^\star = \epsilon^\dagger\Psi^\dagger \label{susytransb}\\ &\delta_\epsilon\Psi_\alpha = -i(\sigma^\mu\epsilon^\dagger)_\alpha \partial_\mu\Phi,~~~~~~~~~~~~~&\delta_\epsilon\Psi^\dagger_{\dot{\alpha}} =  i(\epsilon\sigma^\mu)_{\dot{\alpha}} \partial_\mu\Phi^\star,
   \label{susytransf} 
\end{eqnarray}
and can be written in momentum representation:
\begin{equation}
S = \frac{1}{(2\pi)^4} \int d^4\hat{p}_1 d^4\hat{p}_2~\delta^4 \left( \hat{p}_1+\hat{p}_2 \right) \left[ -\tilde{\Psi}^\dagger(\hat{p}_1)  \bar{\sigma}_\mu \hat{p}_2^\mu \tilde{\Psi}(\hat{p}_2) + \hat{p}_{1\mu} \tilde{\Phi}(\hat{p}_1) \hat{p}_2^\mu \tilde{\Phi}^\dagger(\hat{p}_2)\right],   \label{4dWSmodelp} 
\end{equation}
where $\tilde{\Phi}^\dagger(\hat{p})$ is the Fourier transform of $\Phi^\star(x)$. 
As far as the kinetic terms of Wess-Zumino action are concerned the action 
(\ref{4dWSmodelp}) can be written on the lattice according to the previous 
sections's prescriptions as
\begin{eqnarray}
S^{(z)} &=& \frac{1}{\pi^4} \int_{-\frac{\pi}{a}}^{\frac{\pi}{a}} d^4p_1~d^4p_2~\prod_\mu \delta\left(p_{1\mu} + p_{2\mu} \right)\cdot \nonumber \\ &\cdot& \left[-\tilde{\psi}^\dagger(p_1) \bar{\sigma}^\mu \Delta_G^{(\hat{z})}(p_{2\mu}) \tilde{\psi}(p_2) + 
\Delta_G^{(\hat{z})}(p_{1\mu})\tilde{\varphi}(p_1) \Delta_G^{(\hat{z})}(p_2^\mu)\tilde{\varphi}^\dagger(p_2) \right]. \label{4dWSmodellattpss} 
\end{eqnarray}
Here the regularized momentum operator is used for the 
derivative operators. We can show that this regularized lattice action is 
equivalent to the cut off version of the continuum action:
\begin{equation}
S^{(\hat{z})} = \frac{1}{(2\pi)^4} \int_{|\hat{p}_\mu|\leq \hat{p}^{(\mathrm{cutoff})}} d^4\hat{p}_1 d^4\hat{p}_2~\delta^4 \left( \hat{p}_1+\hat{p}_2 \right) \left[  -\tilde{\Psi}^\dagger(\hat{p}_1)  \bar{\sigma}_\mu \hat{p}_2^\mu \tilde{\Psi}(\hat{p}_2) + \hat{p}_{1\mu} \tilde{\Phi}(\hat{p}_1) \hat{p}_2^\mu \tilde{\Phi}^\dagger(\hat{p}_2)
 \right],   \label{4dWSmodelpct} 
 \end{equation}
with
\begin{equation}
\tilde{\Psi}(\Delta_G^{(\hat{z})}(p_{\mu}))=\frac{2\tilde{\psi}(p)}{\prod_\mu\sqrt{\frac{d\Delta_G^{(\hat{z})}(p_\mu)}{dp_\mu}}}~~~~~~,~~~~~~\tilde{\Phi}(\Delta_G^{(\hat{z})}(p_{\mu}))=\frac{2\tilde{\varphi}(p)}{\prod_\mu\sqrt{\frac{d\Delta_G^{(\hat{z})}(p_\mu)}{dp_\mu}}}. \label{cl} \end{equation}
When the interaction terms are introduced the products are replaced by $\star$-product and the 
procedure goes quite parallel. SUSY is exactly kept on the lattice. 

The action (\ref{4dWSmodelpct}) is the action of a free theory in the continuum where a cutoff on the components of the momenta has been introduced. It is important to notice that in (\ref{4dWSmodelpct}) the dependence on the lattice constant $a$ and the parameter $\hat{z}$, which was explicit in (\ref{4dWSmodellattpss}), is all contained in the value of  $\hat{p}^{(\mathrm{cutoff})}$ which is given in (\ref{bound}). 
As a consequence\emph{ lattice actions  (\ref{4dWSmodellattpss}) with different values of the lattice constant $a$ and of the parameter $\hat{z}$ but corresponding to the same value of the cutoff $\hat{p}^{(\mathrm{cutoff})}$ according to eq. (\ref{bound}) are physically equivalent, as they correspond to the same continuum theory (\ref{4dWSmodelpct}).}

The conventional continuum action is reached by letting $\hat{p}^{(\mathrm{cutoff})} \rightarrow  \infty$. This can be obtained in two ways, namely by either keeping $\hat{z}$ fixed (for instance $\hat{z}=0$) and taking the limit where the lattice spacing $a$ goes to zero (continuum limit), or by keeping the lattice spacing fixed and taking the limit $\hat{z} \rightarrow 1$. The lattice structure is preserved in the limit. 

\subsection{$\Phi^4$ theory in four dimensions}
The lattice field theory formulation that we are proposing is completely general so that we 
can apply it also to non-SUSY field theories. Here we examine $\Phi^4$ theory.   
The momentum representation of $\Phi^4$ theory in four dimensions in the continuum 
is given by
\begin{eqnarray}
S_c &=& \int d^4\hat{p}_1 d^4\hat{p}_2~ \delta^{(4)}(\hat{p}_1+\hat{p}_2) \left[-\hat{p}_1^\mu \tilde{\Phi}(\hat{p}_1)\hat{p}_{2\mu}\tilde{\Phi}(\hat{p}_2) + m_0^2 \tilde{\Phi}(\hat{p}_1)\tilde{\Phi}(\hat{p}_2)\right] \nonumber \\ &+&\lambda_0 \int \prod_{i=1}^4 d^4\hat{p}_i~\delta^{(4)}(\hat{p}_1+\hat{p}_2+\hat{p}_3+\hat{p}_4)~  \tilde{\Phi}(\hat{p}_1)\tilde{\Phi}(\hat{p}_2) \tilde{\Phi}(\hat{p}_3)\tilde{\Phi}(\hat{p}_4), \label{phi4c} 
\end{eqnarray}
The lattice version of this continuum expression can be obtained by replacing derivative 
operators and conserved momenta by the regularized lattice momentum 
$\hat{p}_{i\mu}=\Delta_G^{(\hat{z})}(p_{i\mu})$ and by replacing the continuum wave 
function by its lattice counterpart: 
\begin{equation}
\tilde{\Phi}(\Delta_G^{(\hat{z})}(p_{\mu}))=\frac{2\tilde{\varphi}(p)}{\prod_\mu\sqrt{\frac{d\Delta_G^{(\hat{z})}(p_\mu)}{dp_\mu}}}. 
\label{btl}
\end{equation}
This essentially leads to a cut off theory of $\Phi^4$ theory in the momentum representation. 

The introduction of the regulator $\hat{z}$ in the definition of the lattice derivative 
is not the only possible regularization procedure. Here we propose a different one, 
obtained by modifying the integration volume with the introduction of a parameter 
$\alpha$:  
\begin{equation}
d^4\hat{p}_1\delta^{(4)}(\hat{p}_1+\cdots)\tilde{\Phi}(\hat{p}_1) \rightarrow d^4{p}_1\delta^{(4)}({p}_1+\cdots)
\left(\prod_{\mu=1}^4\frac{d\Delta_G(p_{i\mu})}{dp_{i\mu}}\right)^{2\alpha-1}\tilde{\varphi}(p_1). 
\end{equation}
The momentum representation of lattice $\Phi^4$ action can then be given by 
\begin{eqnarray}
&S^{\alpha}=\int_{-\frac{\pi}{a}}^{\frac{\pi}{a}} d^4p_1 d^4p_2 \prod_{\mu=1}^4 \delta(p_{1\mu}+p_{2\mu}) \left[-\Delta_G(p_{1\mu})\Delta_G(p_{2\mu})+ m_0^2 \left(\prod_\mu \cos \frac{ap_{1\mu}}{2}\right)^{1-2\alpha} \right] \tilde{\varphi}(p_1) \tilde{\varphi}(p_2) \nonumber \\&+\lambda_0  \int_{-\frac{\pi}{a}}^{\frac{\pi}{a}} \prod_{i=1}^4 d^4p_i \prod_{\mu=1}^4\delta\left( \sum_{i=1}^4 \Delta_G(p_{i\mu}) \right) \left( \prod_{i,\mu=1}^4  \cos\frac{ap_{i\mu}}{2} \right)^{-\alpha}  \tilde{\varphi}(p_1) \tilde{\varphi}(p_2) \tilde{\varphi}(p_3) \tilde{\varphi}(p_4),   \label{slatttot} 
\end{eqnarray}

 It is  convenient also in this case to go back to the continuum representation, namely to use $\hat{p}_{i\mu} = \Delta_G(p_{i\mu})$ as independent momenta and $\Phi(\hat{p}_i)$, defined in (\ref{btl}), as fundamental fields.
With this change of variables we find:
\begin{eqnarray}
S^{(\alpha)} &=& \int_{-\infty}^{\infty} d^4\hat{p}_1 d^4\hat{p}_2~ \delta^{(4)}(\hat{p}_1+\hat{p}_2) \left[-\hat{p}_1^\mu \hat{p}_{2\mu} +\frac{m_0^2}{\left(\prod_{\mu} \cosh\frac{a\hat{p}_{1\mu}}{2}\right)^{1-2\alpha}} \right]\tilde{\Phi}(\hat{p}_1)\tilde{\Phi}(\hat{p}_2) \nonumber \\ &+&\lambda_0 \int_{-\infty}^{\infty} \prod_{i=1}^4 d^4\hat{p}_i~\delta^{(4)}(\hat{p}_1+\hat{p}_2+\hat{p}_3+\hat{p}_4)~ \frac{ \tilde{\Phi}(\hat{p}_1)\tilde{\Phi}(\hat{p}_2) \tilde{\Phi}(\hat{p}_3)\tilde{\Phi}(\hat{p}_4)}{\left(\prod_{\mu}\prod_{i=1}^4 \cosh\frac{a\hat{p}_{i\mu}}{2}\right)^{1/2-\alpha}}. \label{phi4alpha} 
\end{eqnarray}
The kinetic term in (\ref{phi4alpha}) is the same as in the standard continuum theory, but the mass and the interaction terms are modified by the presence of the hyperbolic cosine factors which  provide a smooth cutoff in the momenta: in fact for $\alpha<1/2$ each hyperbolic cosine denominator becomes very large for
\begin{equation}
\hat{p}_{i\mu} \gg \frac{1}{a(1/2-\alpha)}. \label{othercutoff} 
\end{equation}

\begin{figure}
	\begin{center}
		\includegraphics[scale=0.5]{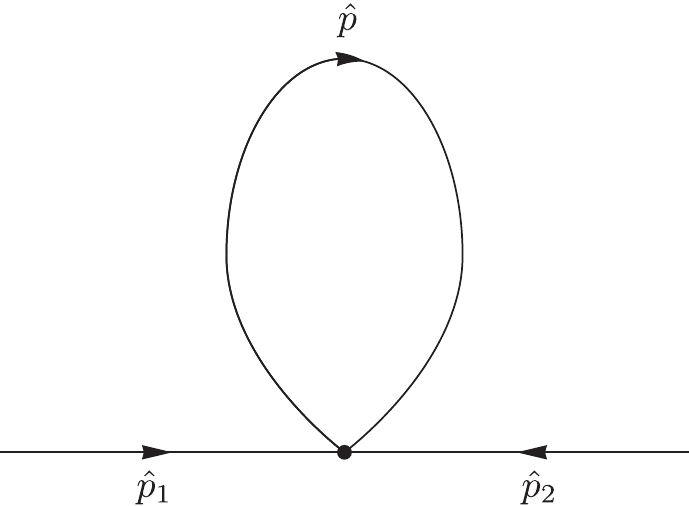}
	\end{center}
	\caption{One-loop correction to the propagator.}
	\label{fd2pt}
\end{figure}

\begin{figure}
	\begin{center}
		\includegraphics[scale=0.5]{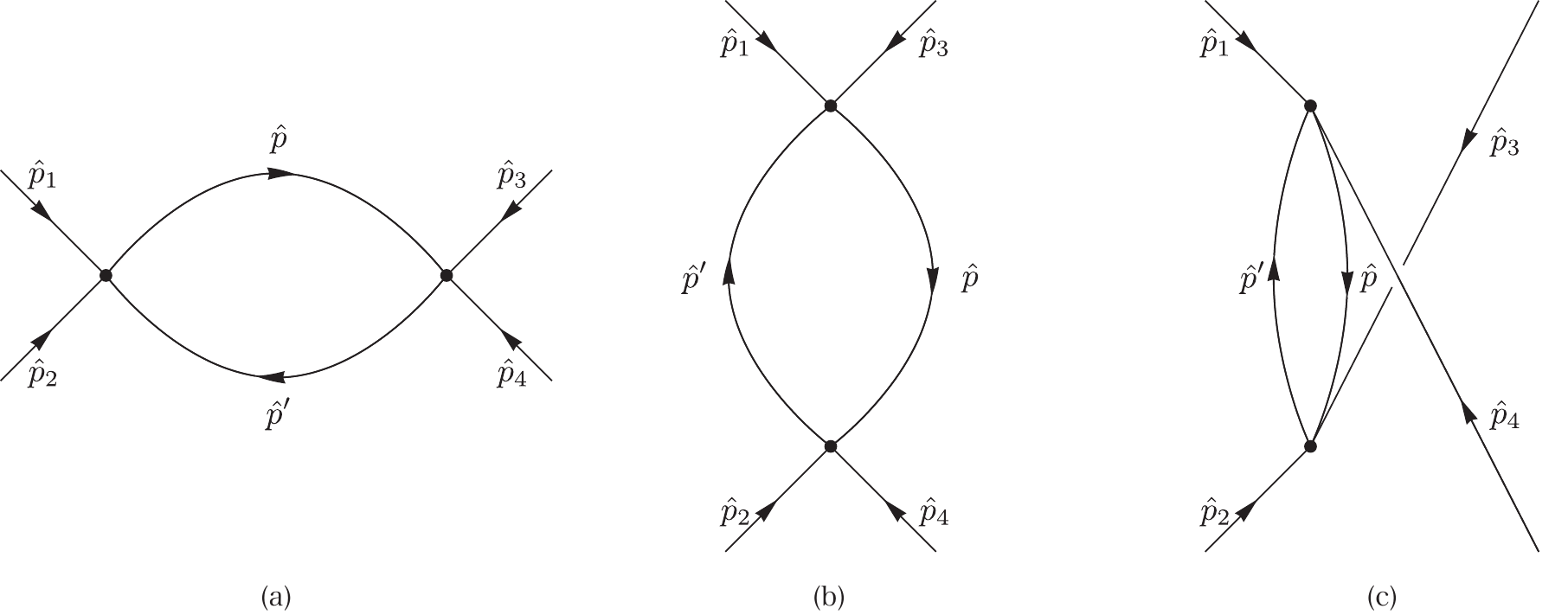}
	\end{center}
	\caption{One-loop corrections to the four point vertex.}
	\label{fd4pts}
\end{figure}

At one loop level the ultraviolet divergent diagrams are the one loop mass renormalization diagram, 
shown in fig.\ref{fd2pt}, 
and the one-loop coupling renormalization diagrams illustrated in fig.\ref{fd4pts}.

The propagator is given by 
\begin{equation}
D^{(\alpha)}(\hat{p}_1,\hat{p}_2) = \frac{\prod_\mu\delta\left(\hat{p}_{1\mu}+\hat{p}_{2\mu}\right)}{\sum_\mu \hat{p}_1^\mu \hat{p}_{1\mu} + \left(\prod_{\mu} \cosh\frac{a\hat{p}_{1\mu}}{2}\right)^{2\alpha-1} m_0^2},  \label{fullprop} 
\end{equation}
which turns into the following after one loop correction:
\begin{equation}
D_{1\textrm{loop}}^{(\alpha)}(\hat{p}_1,\hat{p}_2) = \frac{\prod_\mu\delta\left(\hat{p}_{1\mu}+\hat{p}_{2\mu}\right)}{\sum_\mu\hat{p}_1^\mu \hat{p}_{1\mu} + \left(\prod_{\mu} \cosh\frac{a\hat{p}_{1\mu}}{2}\right)^{2\alpha-1} \left(m_0^2+\lambda_0 I_\alpha\right)}, \label{fullprop1loop} 
\end{equation}
where 
\begin{equation}
I_\alpha = b_2 \left(\frac{1}{a(1/2-\alpha)}\right)^2 + b_0 ~ m_0^2 \log(1/2-\alpha) + \textrm{regular terms},  \label{singIa}. 
\end{equation} 

Similarly coupling constant renormalization can be carried out: 
\begin{equation}
\lambda_0 \longrightarrow \lambda_0 + \lambda_0^2 \left( I_\alpha(\hat{p}_1+\hat{p}_2)+ I_\alpha(\hat{p}_1+\hat{p}_3)+ I_\alpha(\hat{p}_3+\hat{p}_2)\right), \label{lambdacorr} 
\end{equation}
where 
\begin{equation}
 I_\alpha(\hat{p}) = c_0 \log(1/2-\alpha) + \textrm{regular terms}.  \label{singIap} 
\end{equation}

It is interesting to realize that the nonlocal nature of the lattice formulation is 
reflected as the momentum dependent mass term and coupling constant in the 
equivalent continuum theory. Nevertheless we can carry out renormalization 
procedure as in the standard treatment.

\section{Conclusion and Discussions}

We find a lattice formulation which is equivalent to the corresponding continuum 
theory. The problem of the species doubler d.o.f. for chiral fermions is consistently avoided. 
Explicit relations between continuum and lattice transformation 
of fields are derived and can be identified as a block spin type transformation. 
The reversibility of the transformation requires a special choice of lattice translation generator.   
The lattice formulation is nonlocal in nature but it recovers exponential locality 
in the continuum limit. All symmetries including lattice SUSY of the corresponding 
continuum theory can be kept exact, however, the regularized version still lacks associativity 
and thus lacks exact gauge invariance although it is assured to recover in the continuum limit. 

It is worth to mention that the continuum-lattice duality introduced at the classical level by 
the reversible blocking transformation 
can be extended to the quantum level, and the lattice actions obtained in this way may be regarded as 
{\it perfect actions} \cite{Hasenfratz, Bietenholz}, with no lattice artifact in spite of finiteness of $a$. 

It is a natural question if this lattice field theory formulation can be numerically feasible to 
carry out. The lattice field theory is nonlocal but all necessary expressions are well defined. 
Therefore we consider that numerical evaluation is in principle possible. 
The most naive regularization procedure for lattice regularization is to put the system 
in a box. Then the momentum is discretized then vanishing constraints of the 
continuum momentum $\Delta_G(p)$ in the delta function of  (\ref{assstarprodpr}) do not 
have a solution in general\cite{Bergner}. This would cause another source of breaking 
symmetries. We can expect, however, that all the breaking effects would be negligible 
in the continuum limit since in the infinite box limit the lattice formulation is equivalent to 
the continuum theory.  

In this approach lattice gauge theories and super Yang-Mills theories 
are still difficult to formulate in such a way that the gauge symmetry is strictly preserved.  
It could, however, be possible to evaluate the amount of breaking in the process 
of blocking transformation.

\vspace{1cm}
\begin{center}
{\bf{\Large Acknowledgments}}
\end{center}
We thank P. H. Damgaard, I. Kanamori, H. B. Nielsen and Hiroshi Suzuki,  
for useful discussions over long years of our investigations. 
We also wish to thank JSPS and INFN for supporting our long standing collaboration.

\bibliography{lattice2017}

\end{document}